\documentclass[twocolumn,preprintnumbers,aps,superscriptaddress]{revtex4}

\usepackage{amsmath}
\usepackage{amssymb}

\input epsf
\input rotate

\def\bea{\begin{eqnarray}}
\def\eea{\end{eqnarray}}
\def\ben{\begin{equation}}
\def\een{\end{equation}}
\def\benu{\begin{enumerate}}
\def\enu{\end{enumerate}}

\def\1var{(\bx_1...\bx\N)}

\def\b1{{\bf 1}}
\def\bx{{x}}

\def\N{_{\sss N}}

\def\sph_int{ {\int d^3 r}}

\def\infintd3r{ \int_{-\infty}^\infty d^3r\,}
\def\intd3r{ \int d^3r\,}

\def\laplace1d{\frac{d^2}{dx^2}}
\def\plaplace1d{\frac{d^2}{d{x'}^2}}

\def\padr2{\frac{\partial^2}{\partial r^2}}

\def\N{{\cal N}}

\begin{document}

\title{Comment on ``Analysis of Floquet formulation of time-dependent density-functional theory'' [Chem. Phys. Lett. {\bf 433} (2006), 204]}
\author{Neepa T. Maitra}
\affiliation{Department of Physics and Astronomy, Hunter College
and City University of New York, 695 Park Av. New York, NY 10021,
USA.}
\author{Kieron Burke}
\affiliation{1102 Natural Sciences 2 University of California, Irvine, California, 92697-2025}
\date{\today}

\begin{abstract}
We discuss the relationship between modern time-dependent density functional theory and earlier
time-periodic versions, and why the criticisms in a recent paper
(Chem. Phys. Lett. {\bf 433} (2006) 204) of our earlier
analysis (Chem. Phys. Lett. {\bf 359} (2002) 237) are incorrect.

\end{abstract}

\maketitle 

The idea of a formulation of density functional
theory (DFT) applied directly to Floquet states has attracted much attention
over recent years (see e.g. Refs.~\cite{DG82,B81,TC97}). Such a method
would benefit from the favorable system-size scaling of
density-functional approaches as well as the natural treatment of
time-periodic intense field processes that Floquet approaches
provide. Underlying any DFT is a one-to-one
mapping between densities and applied potentials, which depends on both the particle statistics and the particle interaction. In static DFT, this mapping exists only for the ground-state density~\cite{GB04}.
In time-dependent DFT (TDDFT), established by Runge and Gross~\cite{RG84}, this mapping  depends on the initial state.

 However, the Floquet density functional theory (Floquet DFT) proposed
in earlier work~\cite{DG82,TC97} is based on a one-to-one mapping
between densities and potentials, {\it without} initial-state
dependence.  In a recent letter~\cite{MB02} (henceforth MB), we showed
that this mapping does not exist, so that the time-periodic density of
an arbitrary many-electron Floquet state does not uniquely determine
the potential in which it evolves. If analyzed within the framework of
TDDFT, one can construct a one-to-one mapping, but it depends on the
initial state of the system~\cite{MB02}.  This proof does not exclude
the possibility that a mapping might exist for some specified state, a
hope on which the original works were based.

A recent letter~\cite{SH06} (henceforth SH), incorrectly claims that MB overlooked important points and that in
fact Floquet DFT is well-founded and valid.
There are at least four simple errors in SH:

(1) The concept of a ``ground Floquet state'' used in SH implies the
    existence of an adiabatic theorem for Floquet states. This is
    known {\it not} to exist in
    general~\cite{E72,AB71,HKK97,YD70,H92,LEK72,BDH92,DH99}.
 But, even if it did, the subject of MB was TDDFT applied to any Floquet state, under periodic fields of any field strength, weak or
    strong, as formulated and applied in Ref.~\cite{TC97}. SH
    considers only ``ground Floquet states'', and implicitly their work applies only to
    weak fields (see also point (2)). 

(2) The minimal principle for the quasi-energy that SH use
    holds only for weak, off-resonant driving. In particular, it does
    not hold for strong-fields.

(3) Time-dependent DFT (TDDFT), as formulated by Runge and
    Gross~\cite{RG84}, {\it can} be applied to Floquet states, in
    contrast to what is claimed in SH.


(4) The example in MB is valid, and SH's criticism of it is incorrect.

Points (1) and (2) are errors that stem back to the original paper of
Deb and Ghosh~\cite{DG82} where, although the limitations of the proposed
Floquet DFT are acknowledged in their footnotes and references, they
not explicitly discussed.
Ref.~\cite{DG82} was the basis of the Floquet DFT of Telnov and
Chu~\cite{TC97}, where it was however used for strong-field
applications, and for general Floquet-states, i.e. far beyond its regime of
validity.

Before explaining the points in detail, we first remark on a
fundamental problem when attempting to connect Floquet theory with
(TD)DFT~\cite{ref,O74,CCL04}. Floquet states are only guaranteed to
exist~\cite{O74} in systems with a discrete spectrum (which may be infinite or
finite), yet the theorems of density functional theory are based on
the full Hilbert space, including any continuum. Therefore, adapting
any kind of variational theorem in Floquet theory to density
functional theory requires careful inspection, and is likely invalid
for systems which do not have a purely discrete spectrum, i.e. the vast majority of systems to which DFT is applied.  Our paper
MB showed that even when Floquet states {\it do} exist, there
is no one-to-one mapping between their densities and the potentials, as was
assumed in the Floquet DFT's in the earlier literature~\cite{TC97,DG82}.

We now explain points (1)-(4) above in detail. 

(1) The most important conceptual error is that a ``ground Floquet state''
can be uniquely defined by adiabatically tracking the unperturbed
(field-free) ground-state as the time-periodic field is turned
on. But there are significant, and well-recognized, problems with
defining such a state. First, there is no
adiabatic limit when a complete infinite set of basis states is
included~\cite{HKK97,YD70,H92,LEK72,BDH92,DH99}. For
example, in Ref.~\cite{LEK72}, it is stated
``Demonstration of the existance of a set of quasiperiodic solutions
for an adiabatically switched harmonic potential is somewhat
problematic in general'', going on to cite
Refs.~\cite{YD70,E72,AB71}.  
In Ref.~\cite{H92}, the need for
conditions on the ``ineffectiveness of resonances'' is
discussed. 

Essentially the problem stems from having an increasingly
dense spectrum~\cite{BDH92,H92,AHS90,HKK97}, as eigenvalues are
squeezed into a zone of width $\omega$, the driving frequency. There is a weakly avoided crossing near every point in the zone as a function of the strength of the applied periodic potential, $\lambda$. Quoting from
Ref.~\cite{HKK97}, ``the structure of the exact states and quasienergy
spectrum is remarkably irregular....the familiar quasienergy
"dispersion" curves as functions of $\lambda$...become discontinuous
everywhere. One consequence...is the absence of a true adiabatic
limit; there is no unique final state to which the system tends as the
periodic perturbation is switched on arbitrarily slowly.'' 
In summary,  there is in general no adiabatic
limit for Floquet states within a complete infinite Hilbert
space. 

Several works have nevertheless derived types of modified adiabatic
theorems for Floquet states, but each require some further assumption
or approximation~\cite{AHS90,HKK97,YD70,DH99}. Often truncation to a
finite basis and studies of convergence of Floquet states with respect
to basis size are made.  Ref.~\cite{HKK97} argue that the effects of
interactions with the environment are likely to restore an adiabatic
theorem for open systems. For these reasons, the usual Floquet methods
apply in many physical situations.



Finally, there is ambiguity in SH regarding how their ``ground Floquet state'' is defined: for example, shortly after
Eq.~(7) in SH, is the statement that ``Here the ``ground state''
refers to a steady-state having the lowest quasienergy''.  But the ``lowest'' depends on the
choice of zone; in this definition, {\it any} state may be chosen as
the ``ground state'' by simply shifting the zone boundary. The
quasi-energies are defined modulo $\omega$, the driving frequency, so
may be chosen to lie in the zone $[x -\omega/2,x+\omega/2)$, where $x$
is any real number~\cite{S73}. In other parts of SH, however, the
ground Floquet state is defined as that obtained by adiabatically
ramping up the field, beginning in the unperturbed static
ground-state.

(2) SH argue that a one-to-one density-potential mapping holds, based
on an energy minimum principle~\cite{LEK72,DG82}.  However,
use of a minimum principle implies existence of a complete set of
Floquet states but, as discussed above point (1), their existence is
called into question when the spectrum of the system possesses a
continuum component. Even if there is no continuum, the proofs of the minimum
principle~\cite{LEK72} depend on adiabatic turn-on, and so hold only
for a basis truncated to a finite number of dimensions.  Yet,
even if we now restrict to systems with a purely discrete finite spectrum,
the minimum principle holds only when the driving frequency $\omega$
``is chosen to insure transitions to excited states cannot
occur''~\cite{LEK72}. For linear response, this means that $\omega$
must not be one of the resonant frequencies of the system. For
increasingly intense fields, this means that the minimum principle
holds only for an increasingly small frequency-region (as explained in
footnote 80 of Ref.~\cite{LEK72}). A
one-to-one density-potential mapping for Floquet states adiabatically
ramped from the unperturbed ground-state has been argued to exist~\cite{DG82} only under the
following approximations: (a) truncation of the problem to a finite
basis, {\it and} (b) only for weak, off-resonant driving. Point (b)
appears to be recognized in some places in SH for linear response, but
on the other hand, SH do not discuss the severe restriction that this
imposes for strong fields, making it inapplicable to Ref.~\cite{TC97}.


One may ask whether the problems of adiabatic ramping may be bypassed
by simply choosing a zone for the quasi-energies and considering a
minimal principle based on the lowest quasi-energy in that zone. But
this approach would require a ``zone-dependence'' in any functional; if
the zone is shifted even slightly, the state with the lowest
quasi-energy may be completely different. Moreover, if the field
strength is altered even slightly, the ``minimal'' state may hop from
the bottom of the chosen zone to the top, and another completely
different state be minimal.

(3) None of these problems occur within the full TDDFT framework of Runge-Gross (RG)~\cite{RG84}, as used in MB. There the one-to-one
mapping depends on the initial-state of the system. The initial time
may be chosen to be any time, in particular, once the time-periodic
Floquet state is established.  SH correctly point out adiabatic
turn-on is excluded from the RG theorem, but incorrectly deduce that this invalidates the use of TDDFT for Floquet phenomena. SH incorrectly state that ``the steady-state solutions are
obtained by an adiabatic switching of the periodic potential''. This
is not true: Floquet states are defined as quasi-periodic solutions to
the Schr\"odinger equation for time-periodic potentials~\cite{S73},
completely independent of any adiabatic switching. This is particularly important in light of point (1).  When one applies RG
TDDFT to a Floquet state, one assumes the system is already in the
Floquet state. As in TDDFT applied to a general problem, the initial
time and initial state may be chosen to be any time at which the
interacting and Kohn-Sham wavefunctions are known~\cite{RG84,MBW02}. But the initial-state dependence of RG implies that the density functionals are different from those that are almost always used in TDDFT applications today, i.e. when the system starts from a ground state. 


(4) We now turn to the discussion in SH of the example of MB. The authors
claim that ``...Maitra and Burke are incorrect on two counts. First,
they consider a system in steady-state solution that is supposed to
have been obtained by an adiabatic switching.'' We never
state this and, as discussed above, one cannot
assume adiabatic switching in the general case without further assumptions. SH
concludes then that RG  is not
applicable to this system, but this is incorrect as explained in point (3).
SH devote a long discussion to the similarity of Floquet states with
excited states in time-independent problems, and the problems with
uniqueness of mappings for excited states, claiming MB ``fail to make'' this connection. This is however
well-recognized in several works~(e.g. \cite{GB04,MB01,MBW02, G99}) containing explicit examples. 
Indeed one of the examples discussed in
detail in SH is, up to a trivial change in parameter, identical to that of the uncited Ref.~\cite{MBW02}!

We close by noting that the example of the periodically driven
harmonic oscillator in MB is an exceptional case from the point of
view of adiabatic turn-on, since the quasi-energies monotonically
increase as a function of the driving strength, but with
discontinuities arising solely from each quasi-energy being knocked
down from the top to the bottom of the chosen quasi-energy-zone. 
For this special case, it may be argued
that the state without spatial nodes could be called a "ground-Floquet state". 
This
example is nevertheless used correctly in MB to simply show the non-uniqueness 
property, i.e. that one may find different Floquet
states that evolve with the same time-periodic density in different
time-periodic potentials. It is clear that examples may be constructed
in the same way for more generic potentials whose quasi-energy spectra
display the more typical discontinuities discussed in
Ref.~\cite{HKK97}. e.g., ``kicked rotor'' (driven free-particle in a
box).

Finally, we note that within the approximate finite-basis methods implicit
in the numerical treatment of many linear response approaches, as in
Refs.~\cite{SHS04,KA05} a Floquet approach is redeemed, as the
problems discussed above are bypassed.

NTM is financially supported by an NSF Career grant, CHE-0547913, and the Hunter College Gender Equity Project. KB is supported by NSF CHE-0355405.

. 

\end{document}